# Bulk superconductivity in quintuple-layer square-planar nickelate $La_6Ni_5O_{12}$

Mengzhu Shi[1#], Mei Du[1#], Hongyu Li[1], Zhiwei Wang[2], Zhengjie Wang[2], Ziji Xiang[1,4], Tao Wu[1,2,3,4,*], Xianhui Chen[1,2,3,4,*]

[1] Hefei National Research Center for Physical Sciences at the Microscale, University of Science and Technology of China, Hefei 230026, China

[2] Department of Physics, University of Science and Technology of China, Hefei 230026, China

[3] Collaborative Innovation Center of Advanced Microstructures, Nanjing University, Nanjing 210093, China

[4] Hefei National Laboratory, University of Science and Technology of China, Hefei 230088, China.

[#]These authors contributed equally: Mengzhu Shi and Mei Du.

*Correspondence to: wutao@ustc.edu.cn, chenxh@ustc.edu.cn

**Nickelate superconductors, sharing a similar electronic environment with cuprate superconductors, offer a promising platform for unraveling the mechanism of high-temperature superconductivity. However, the realization of superconductivity in these systems has so far been restricted to thin films or bulk samples under high-pressure. This limits the applicability of various experimental probes to study the mechanism of superconductivity, and hinders the potential application of the nickelate superconductors. Exploring bulk superconductivity at ambient pressure is therefore essential to understand its origin and enable broader experimental research. Here, we report the realization of bulk superconductivity at ambient-pressure in quintuple-layer square-planar (SP) nickelate $La_6Ni_5O_{12}$. The SP phase is obtained by performing topotactic reaction on the quintuple-layer Ruddlesden-Popper (RP) compound $La_6Ni_5O_{16}$. Optimized reduction condition endows $La_6Ni_5O_{12}$ with a superconducting state that exhibits onset transition temperature of ~ 18 K, as verified by both zero resistance and diamagnetic shielding effect. Pronounced hysteresis loops revealed by magnetization and torque magnetometry demonstrate the occurrence of vortex pinning. Specific heat measurements on a polycrystalline pellet of $La_6Ni_5O_{12}$ reveal a clear superconducting transition at ~18 K and a superconducting volume fraction of ~ 25%, which strongly**

**supports the presence of bulk superconductivity. Our results establish an unprecedented bulk sample platform for investigating superconductivity in nickelates at ambient pressure, which paves the way for further experimental studies into the superconducting mechanism.**

## Introduction

High-temperature superconductivity has remained a central frontier of condensed matter physics since the discovery of cuprate superconductors [1]. To date, cuprate oxides are the only family of unconventional superconductors that sustain bulk superconductivity above the liquid-nitrogen temperature (77 K) at ambient pressure. Despite extensive theoretical and experimental efforts, the mechanism governing high-temperature superconductivity in these strongly correlated systems remains elusive [2]. This long-standing puzzle motivates the exploration of structurally and electronically analogous materials, which may provide crucial insights into the origin of high-temperature superconductivity [3].

Nickel-based oxides have long been regarded as promising analogs of cuprates owing to their structural and electronic similarities [3, 4]. A major breakthrough occurred in 2019 with the discovery of superconductivity in infinite-layer nickelate thin films ($Nd_{1-x}Sr_xNiO_2$) [5], which experimentally established that nickel oxides can host superconductivity [6–8]. Subsequently, superconductivity has been discovered in multilayer nickelates, including square-planar (SP) and Ruddlesden–Popper (RP) phase nickelates [9–16]. So far, however, all nickelate superconductivity has been restricted to thin-film architectures or high-pressure environments, and bulk superconductivity at ambient pressure has not yet been achieved. This limitation restricts the range of experimental techniques available for in-depth studies of the superconducting mechanism in nickelates.

To overcome the limitations inherent to infinite-layer nickelate thin films, extensive efforts have been made to realize superconductivity in bulk single crystals of the same structural family, but all such attempts have so far been unsuccessful [17–21]. Transport measurements on bulk $Nd_{1-x}Sr_xNiO_2$ prepared via topochemical reduction consistently exhibit insulating or semiconducting behavior, with no trace of superconductivity even under applied pressures up to 50.2 GPa [17, 18]. Systematic scanning transmission electron microscopy (STEM) studies indicate that the topotactic reaction produces three-dimensional (3D) block-like structural domains, which disrupt intragrain continuity and induce microcracks. This structural disorder is likely a key factor responsible for the insulating character and for suppressing superconductivity in bulk infinite-layer nickelates [22]. Although the metallicity of infinite-layer nickelate single crystals has been substantially improved, no superconductivity has been observed to date [20, 21].

On the other hand, multilayer SP-phase nickelates have also been predicted to exhibit cuprate-like superconductivity [23, 24]. Previous studies suggest that the fluorite spacer layer in multilayer SP-phase nickelates can stabilize the compound against microcrack formation during topotactic reduction [25]. In particular, the discovery of superconductivity in $Nd_6Ni_5O_{12}$ thin films, together with first-principles calculations showing that the n = 5 member naturally lies in a cuprate-like doping regime, has made the exploration of superconductivity in bulk quintuple-layer nickelate crystals a compelling target [9, 14, 26–30]. However, bulk quintuple-layer nickelates have not yet been successfully synthesized through conventional routes.

Here we report the first high-temperature, high-pressure synthesis of the quintuple-layer RP nickelate $La_6Ni_5O_{16}$ in bulk form. Subsequent topotactic reduction yields ambient-pressure bulk superconductivity in the quintuple-layer SP nickelate $La_6Ni_5O_{12}$, with a superconducting onset temperature of approximately 18 K. This represents the first realization of bulk superconductivity in nickelate superconductors at ambient pressure.

**Quintuple-layer nickelates with Ruddlesden-Popper and Square-Planar structure**

We successfully synthesized the quintuple-layer nickelate $La_6Ni_5O_{16}$ in bulk form for the first time, using a high-temperature, high-pressure six-anvil technique (see Methods for details). Single-crystal X-ray diffraction data collected on selected crystals establish that the as-grown phase crystallizes in a monoclinic Ruddlesden–Popper structure, comprising five perovskite $(LaNiO_3)_5$ sublayers separated by rock-salt LaO layers (Fig. 1a and Extended Data Table 1). This structural model is directly corroborated by HAADF-STEM and iDPC-STEM imaging, which resolve the atomic stacking sequence and reproduce the arrangement expected from the single-crystal diffraction refinement (Fig. 1b, c and Extended Data Fig. 1).

Topotactic reduction induces a pronounced structural reconstruction. The resulting phase is indexed as tetragonal T′-type $La_6Ni_5O_{12}$, consisting of five infinite-layer $NiO_2$ planes alternating with fluorite-type blocking layers (Fig. 1a). This transformation originates from the removal of apical oxygen atoms from the $NiO_6$ octahedra, thereby converting the perovskite-derived nickelate layers into square-planar $NiO_2$ planes analogous to those in cuprate superconductors. HAADF-STEM and iDPC-STEM images of the annealed crystals resolve the transformed stacking sequence and show excellent agreement with the structure determined by single-crystal X-ray diffraction (Fig. 1d, e and Extended Data Fig. 2). Similar structural conversions have previously been reported in other Ruddlesden–Popper phase nickelates with n = 2 ($La_3Ni_2O_7$) and n = 3 ($La_4Ni_3O_{10}$) [31, 32].

Powder X-ray diffraction further confirms the phase conversion (Fig.1f, g and Extended Data Fig.3). The diffraction pattern of the as-grown product is dominated by reflections from the $La_6Ni_5O_{16}$ phase, and its peak positions agree well with those calculated from the single-crystal structural model (Fig. 1f and Extended Data Table 2 ). After topotactic reduction at 320 °C for 40 h, the measured powder diffraction pattern matches the calculated reflections of tetragonal $La_6Ni_5O_{12}$ (Fig. 1g and Extended Data Table 3), with a small fraction of incompletely deoxygenated (partially reduced) $La_6Ni_5O_{16}$ coexisting in the powder sample. The corresponding Rietveld refinement parameters are listed in Extended Data Table 2 and 3. Collectively, the diffraction and atomic-resolution microscopy results establish the conversion of monoclinic $La_6Ni_5O_{16}$ into tetragonal $La_6Ni_5O_{12}$. We note the presence of small cracks on the surface of the annealed product (see inset of Fig. 1g). Charge-neutrality considerations yield a nominal Ni valence of +1.2 in $La_6Ni_5O_{12}$, matching that of optimally doped 112-type infinite-layer nickelate superconductors. A more detailed discussion of the oxygen content analysis will be presented in the next section.

**Superconductivity at 18 K within a limited topotactic reduction window**

We next examined the evolution of electronic transport as a function of topotactic reaction temperature. It should be noted that all transport measurements are performed on selected microcrystals (see Methods for details). The as-grown $La_6Ni_5O_{16}$ sample displays metallic resistivity over the measured temperature range, without a resolvable density-wave anomaly (Fig. 2a). Reduction at 310 °C preserves metallic transport at high temperatures but introduces two low-temperature features: a pronounced resistive anomaly near 94 K, followed by a sharp decrease in resistivity below approximately 18 K. The anomaly near 94 K exhibits clear hysteresis between cooling and warming cycles (Fig. 2b), indicative of a first-order transition. While its microscopic origin remains elusive, its emergence in the incompletely reduced regime is consistent with a density-wave-like instability driven by residual apical oxygen or local oxygen inhomogeneity resulting from partial reduction of RP structure [33,34]. It should be addressed that the applied magnetic fields up to 9 Tesla along the *c*-axis have no effect on the anomaly around 94 K as shown in Fig.2b.

The low-temperature resistive transition is systematically suppressed by a magnetic field applied along the $c$-axis (Fig. 2c). This field response identifies the transition as superconducting. Fitting the onset transition fields using the empirical Ginzburg–Landau expression (see Methods for more details) yields $\mu_0 H_{c2} \approx 34.2$ T for the 310 °C-annealed crystal (Fig. 2d). We note that a small non-zero resistance persists at low temperatures, which is likely attributable to residual apical oxygen, or local oxygen inhomogeneity or the cracks shown in the morphology of the inset of Fig.1g. Annealing at 320 °C yields a metallic normal state followed by a superconducting transition with an onset temperature of approximately 18 K (Fig. 2a, e). The superconducting transition shifts monotonically to lower temperatures under $H \parallel c$, and the corresponding empirical Ginzburg–Landau fit gives $\mu_0 H_{c2} \approx 41.5$ T (Fig. 2f, g). It should be pointed out that a small non-zero resistance value at low temperatures is often observed even in the 320 °C-annealed samples, which is most likely due to massive cracks formed during the topotactic reaction process (see the morphology shown in the inset of Fig. 1g).

Further increases in topotactic reaction temperature to 330 °C drive the samples into an insulating transport regime (see inset of Fig. 2a). The sharp evolution from a coexistence of density-wave state and superconducting state at insufficient reduction, through an optimal superconducting regime, to an insulating state at stronger reduction highlights the exceptional sensitivity of $La_6Ni_5O_{12}$ to oxygen stoichiometry and the resulting in sensitive change of Ni valence. Incomplete removal of apical oxygen from the RP structure likely drives the density-wave-like anomaly as similar as that in $La_3Ni_2O_7$ [33] and $La_4Ni_3O_{10}$ [34]. However, over-reduction risks introducing planar oxygen defects that suppress metallic transport, similar to the behavior of bulk 112 materials [17]. Thus, clean superconductivity is stabilized within a narrow annealing-temperature window.

Furthermore, we have measured the average oxygen content of the powder product by thermogravimetric analysis (see Method for more details). For as-grown $La_6Ni_5O_{16}$ sample, the determined average oxygen content is 15.9. For $La_6Ni_5O_{12}$ powder samples under different topotactic reduction temperatures, the determined average oxygen contents are 13.1 (310 °C), 12.7 (320 °C), and 12.2 (330 °C), respectively. The powder samples may contain the phases with inhomogeneous oxygen contents, so that the average oxygen content obtained from bulk TGA is likely slightly higher than the intrinsic oxygen content of the $La_6Ni_5O_{12}$ phase. Qualitatively speaking, these results are consistent with above discussion on the evolution of resistivity behavior due to oxygen reduction. More precise determination of the oxygen stoichiometry for the $La_6Ni_5O_{12}$ single crystals will require future microscale characterization, such as multislice electron ptychography [35].

**Magnetic evidences for superconductivity in $La_6Ni_5O_{12}$ single crystals**

We further established the superconducting nature of $La_6Ni_5O_{12}$ reduced at 320 °C through magnetic measurements. Figure 3a shows the zero-field-cooled (ZFC) and field-cooled (FC) magnetic susceptibilities measured under 100 Oe on a set of seven single crystal pieces with a total mass of ≈ 0.287 mg. A diamagnetic response emerges at approximately 18 K, consistent with the resistivity transition. The superconducting shielding fraction is estimated to approach 26%; this fraction is likely limited by the inhomogeneity of oxygen content.

The field-dependent magnetization at 2 K displays a clear hysteresis loop (Fig. 3b), characteristic of irreversible vortex dynamics in a type-II superconductor. From the initial low-field response, we estimate a lower critical field of approximately 950 Oe. Applying a spherical-sample approximation to the loop width yields a critical current density of about 2000 A $cm^{-2}$ (see Methods for details). This value demonstrates appreciable current-carrying capability in the $La_6Ni_5O_{12}$ crystals, which is comparable to that of $La_2PrNi_2O_7$ films [36] and supports a bulk superconductivity.

To circumvent the issue of large-scale oxygen inhomogeneity, we utilized magnetic torque, a highly sensitive magnetization probe, to investigate the superconducting magnetic response in individual single crystals. Measurements of the magnetic torque $\tau(H, T)$ on an individual crystal further verify the occurrence of bulk superconductivity. Under an applied field $H$ of 0.5 T, the $\tau(T)$ curve changes sharply near 18 K (Fig. 3c), coincident with the transition temperatures obtained from susceptibility and transport measurements. Moreover, the $\tau(H)$ curve measured at 2 K exhibits a pronounced hysteresis loop (inset, Fig. 3c); this is a clear manifestation of flux pinning in the superconducting state, in accord with the hysteretic $M(H)$ discussed above. The hysteresis loop of transverse magnetization $M_{\perp}(H) = \tau(H)/\mu_0 H$ (Fig. 3d). gradually diminishes upon warming and becomes significantly suppressed at 25 K. Since the diamagnetic shielding effect and the evolution of hysteretic $M_{\perp}(H)$ and $M(H)$ consistently point towards the same transition temperature of $T_c \approx 18$ K, we conclude that the bulk superconductivity developed below this temperature in $La_6Ni_5O_{12}$ single crystals. We note that samples annealed at 310 °C also exhibit similar superconductivity in torque measurements (see Extended Data Fig. 7).

**Evidence for bulk superconductivity from specific heat**

Bulk superconductivity in $La_6Ni_5O_{12}$ obtained by annealing at 320 °C was further confirmed by the measurement of specific heat $C_p$ (see Methods for more details). As displayed in Fig. 4a, the unprocessed zero-$H$ data of $C_p(T)/T$ unambiguously reveal an anomaly at approximately 18 K, which shifts to lower $T$ and is suppressed with increasing magnetic fields. The location of this anomaly congruently matches the superconducting transition determined from magnetic properties (Figs. 3a and c); the occurrence of such a superconductivity-induced jump anomaly in $C_p$ represents compelling thermodynamic evidence that $La_6Ni_5O_{12}$ is a bulk nickelate superconductor. We isolated the electronic specific heat $C_e(T)$ by subtracting the phononic contribution $C_{ph}$ from the zero-$H$ $C_p(T)$ (Methods). Our background analysis is validated by the fact that a fit incorporating both $C_{ph}$ and a normal-state $C_e(T) = \gamma_n T$ (red curve in Fig. 4b, $\gamma_n$ is the Sommerfeld coefficient in the normal state) satisfyingly traces the $C_p(T)$ data taken at $\mu_0 H$ = 14 T where the jump anomaly is almost suppressed, see Fig. 4b. The obtained zero-$H$ $C_e(T)/T$ (Fig. 4c) harbors a residual zero-$T$ term $\gamma_0$ arising from the non-superconducting part of the sample; $\gamma_0$ contributes to approximately 75% of the normal-state $\gamma_n$ estimated for $T > T_c$. This implies a molar fraction of ≈ 25% of the superconducting composition in our polycrystalline $La_6Ni_5O_{12}$ sample, in agreement with the result of diamagnetic volume shielding fraction (Fig. 3a). Moreover, the superconducting component of $C_e(T)$ is well described by a single-band BCS $\alpha$-model [37, 38]. A best fit to this model (solid line in Fig. 4c) gives a mean-field transition temperature $T_c$ = 17.5 K and a superconducting gap $\Delta$ = 3.6 meV. This result of $T_c$ is reinforced by the analysis of entropy balance [39] (see Methods for more details) that yields a very close value of 17.8 K , as shown in the inset of Fig. 4c. We also remark that there is a good consistency among $T_c$ determined from different experimental probes.

## Discussion and Conclusion

Here, we demonstrate that ambient-pressure bulk superconductivity at $T_c$ = 18 K can be achieved in quintuple-layer square-planar nickelate $La_6Ni_5O_{12}$ with a nominal Ni valence of +1.2. This valence state is comparable to that found in superconducting $Nd_{1-x}Sr_xNiO_2$ thin films. A critical aspect of this discovery lies in the electronic structure of the n = 5 member. Previous theoretical calculations have suggested that quintuple-layer nickelates possess an electronic structure highly analogous to that of infinite-layer nickelates [26-30]. Specifically, the multi-layer $NiO_2$ planes in $La_6Ni_5O_{12}$ are expected to host Ni $3d_{x^2-y^2}$ orbitals that form the primary conduction bands near the Fermi level, similar to the single-band physics often discussed in the context of cuprates [3, 4]. The realization of superconductivity in bulk $La_6Ni_5O_{12}$ crystal supports the theoretical prediction that the n = 5 system naturally resides in a doping regime favorable for superconductivity [24].

Furthermore, the availability of ambient-pressure superconducting single crystals eliminates many obstacles for bulk probes. These crystals provide sufficient volume for neutron scattering and nuclear magnetic resonance measurements, naturally cleavable surfaces for scanning tunneling microscopy and angle resolved photoemission spectroscopy, and a substrate-free environment that allows the intrinsic bulk electronic structure to be probed without interfacial or strain-related ambiguities.

In conclusion, we have successfully synthesized quintuple-layer square-planar nickelate $La_6Ni_5O_{12}$ and observed bulk superconductivity at ~ 18 K with the superconducting volume fraction of ~ 25%. This work establishes an unprecedented bulk sample platform for investigating superconductivity in square-planar nickelates at ambient pressure. It paves the way for further experimental studies into the superconducting mechanism, particularly regarding the role of multi-layer electronic correlations and the comparison with infinite-layer thin films. This experimental platform also creates a realistic path toward the systematic optimization of superconducting properties through chemical substitution, annealing protocols, and structural engineering, encouraging the search for additional ambient-pressure nickelate superconductors in the broader RP phase series.

## Main text figures and tables

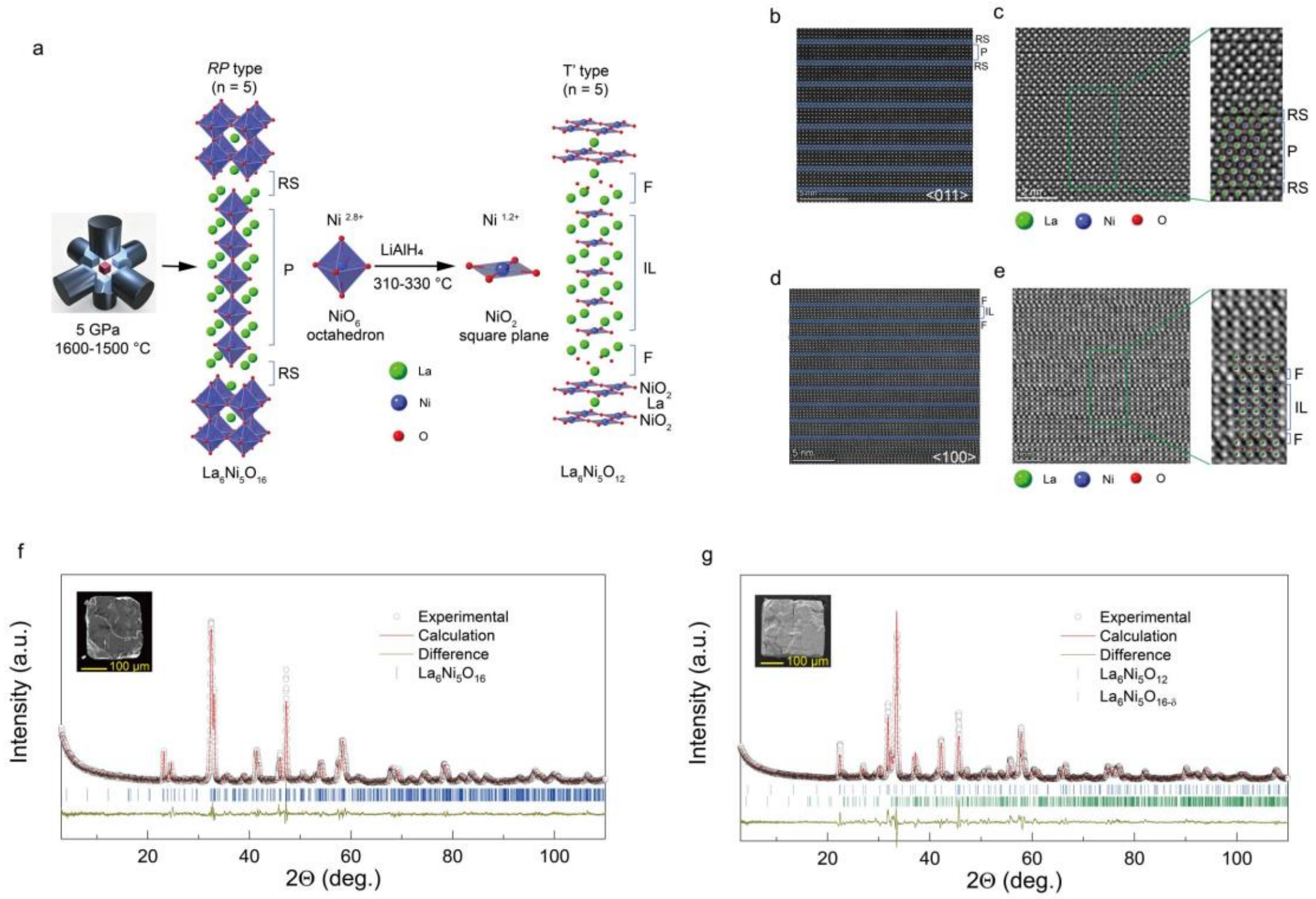

**Fig. 1 | Quintuple-layer nickelates with Ruddlesden-Popper and Square-Planar structure.** **a**, Schematic illustration of the structural transformation induced by topotactic reduction. High-pressure, high-temperature synthesized single-crystalline $La_6Ni_5O_{16}$ crystallizes in a monoclinic Ruddlesden–Popper (RP) structure, comprising alternating five-layer perovskite (P) $(LaNiO_3)_5$ sub-blocks and rock-salt (RS) LaO sub-layers. The removal of apical oxygen atoms from the $NiO_6$ octahedra yields tetragonal T'-type $La_6Ni_5O_{12}$, in which five infinite-layer $NiO_2$ (IL) planes alternate with fluorite-type (F) sub-blocks. **b**, HAADF-STEM image of $La_6Ni_5O_{16}$ taken along the <011> zone axis, showing the five-layer perovskite blocks separated by rock-salt $La_2O_2$ layers. **c**, iDPC-STEM image of $La_6Ni_5O_{16}$ taken along the <011> zone axis (left), in which the five-layer perovskite blocks separated by $La_2O_2$ layers are clearly visible and the oxygen atoms can be directly resolved; the right panel is an enlarged view of the boxed region, which matches well with the structure obtained from the four-circle single-crystal X-ray diffraction refinement. **d**, HAADF-STEM image of $La_6Ni_5O_{12}$ taken along the <100> zone axis, showing the five $NiO_2$ planes separated by $La_2O_2$ layers. **e**, iDPC-STEM image of $La_6Ni_5O_{12}$ taken along the <100> zone axis (left), showing the five $NiO_2$ planes separated by fluorite-type $La_2O_2$ layers, with the apical oxygen atoms between adjacent $NiO_2$ planes completely removed. **f**, Rietveld refinement of the powder XRD pattern of $La_6Ni_5O_{16}$ (powder ground from the entire reaction product), showing the observed data (black circles), calculated

fit (red solid line), Bragg positions of the target phase (green vertical ticks), and the difference curve (blue line, offset below). The inset is an optical micrograph of the small crystals picked from the parent product. **g**, Rietveld refinement of the powder XRD pattern of the 320 °C-annealed sample (powder ground from the entire reaction product), showing the observed data (black circles), calculated fit (red solid line), Bragg positions of the target phase (green vertical ticks), and the difference curve (blue line, offset below); a fraction of an incompletely deoxygenated product coexists with the target phase. The inset is an optical micrograph of the small crystals picked from the 320 °C-annealed product, where shows many cracks on the surface of the crystals..

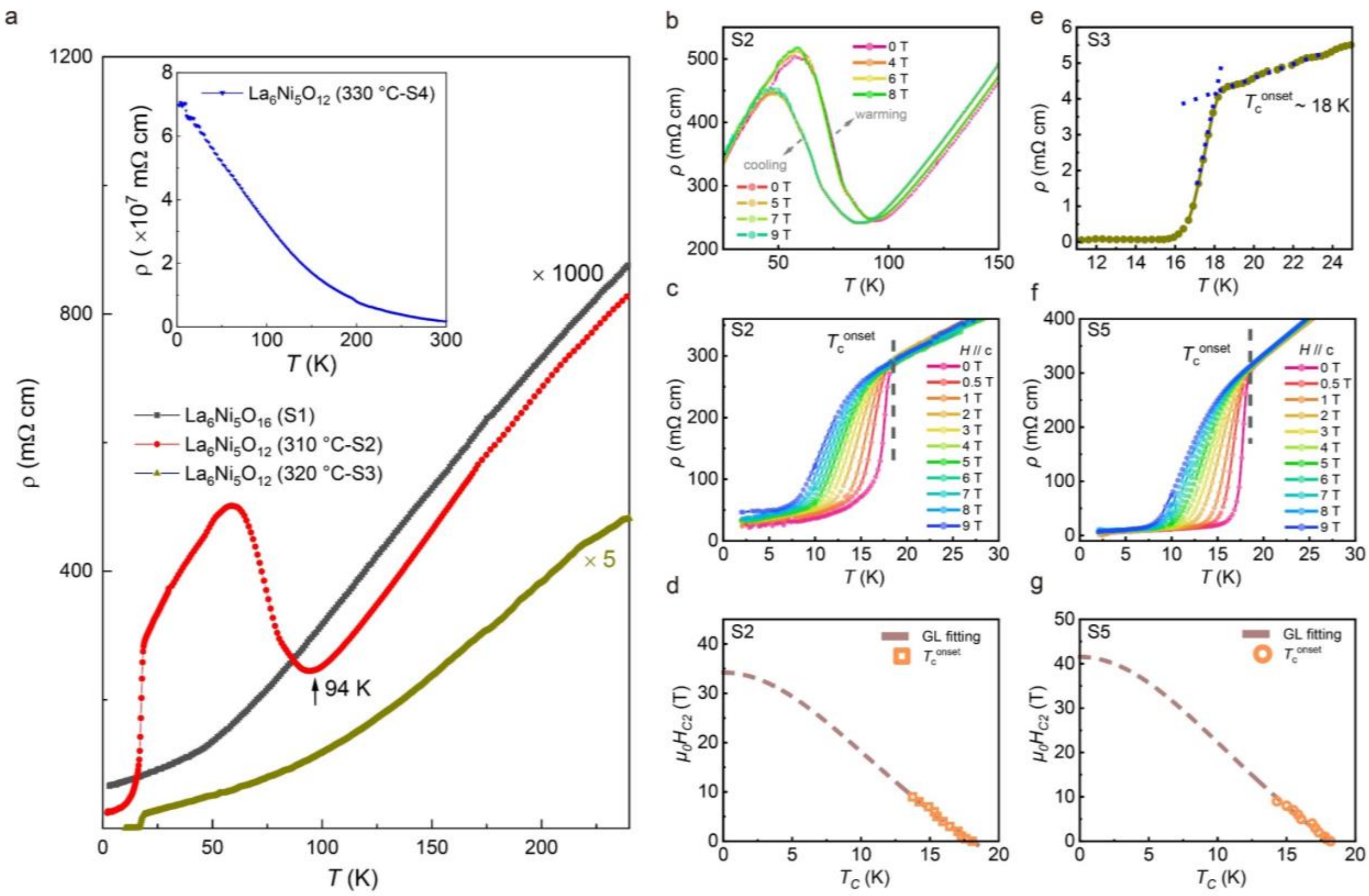

**Fig. 2 | Superconductivity in $La_6Ni_5O_{12}$ realized within a limited hydrogen annealing window**. **a**, Temperature-dependent resistivity of as-grown $La_6Ni_5O_{16}$ and crystals annealed in $H_2$ at 310, 320, and 330 °C, respectively. The as-grown $La_6Ni_5O_{16}$ crystal exhibits metallic transport without any discernible density-wave transition. Superconducting transitions emerge for the samples annealing at 310 and 320 °C, whereas the sample S4 annealed at 330 °C (inset) displays insulating behavior. **b**, Resistivity measured during cooling and warming cycles for the sample S2 annealed at 310 °C. The anomaly near 94 K exhibits pronounced thermal hysteresis, consistent with a first-order density-wave-like transition, and no effect on the transition by applied magnetic field up to 9 Tesla. **c**, Low-temperature resistivity of the sample S2 annealed at 310 °C under magnetic fields applied along the *c*-axis (*H*//*c*). The superconducting transition around 18 K shifts progressively to lower temperatures with increasing field. **d**, Temperature dependence of the upper critical field, $\mu_0 H_{c2}$, determined from the onset transition temperatures for the sample S2 annealed at 310 °C. The dashed curve represents a Ginzburg–Landau fit, yielding a $\mu_0 H_{c2}$ of approximately 34.2 T. **e**, Superconducting transition at 18 K with zero-resistance for the sample S3 annealed at 320 °C. **f**, Magnetic-field dependence of superconducting transition for the sample S5 annealed at 320 °C. The onset transition occurs near 18 K and is progressively suppressed by *H*//*c*. **g,** A Ginzburg–Landau analysis (see Methods for details) yields a $\mu_0 H_{c2}$ of approximately 41.5 T for the sample S5 annealed at 320 °C.

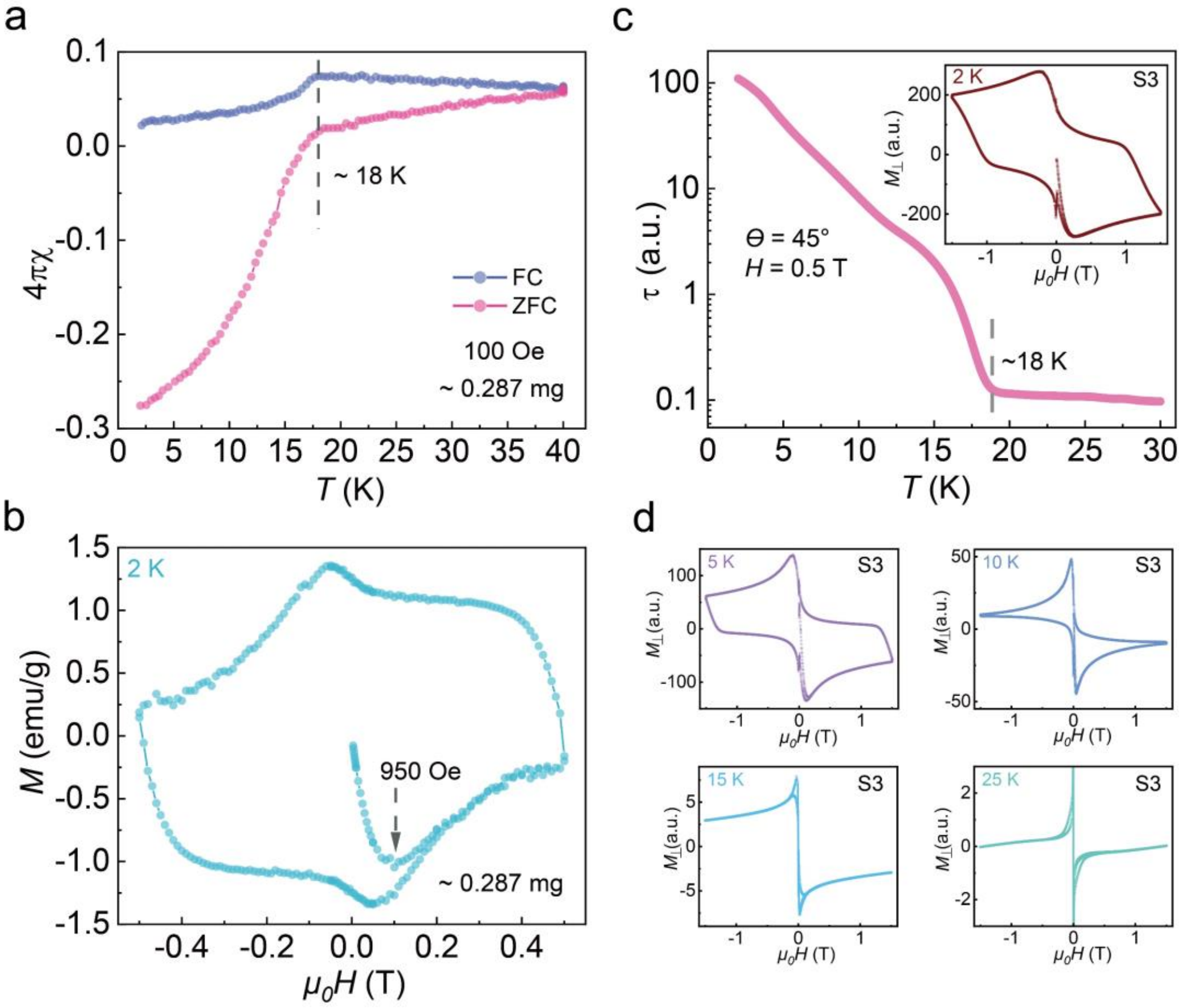

**Fig. 3 | Bulk magnetic signatures of superconductivity in the $La_6Ni_5O_{12}$ crystals annealed at 320 ℃.** **a,** Zero-field-cooled (ZFC) and field-cooled (FC) magnetic susceptibility measured under an applied field of 100 Oe. Data were collected from an ensemble of seven single crystals with a total mass of approximately 0.287 mg. A diamagnetic transition onset near 18 K, with an estimated superconducting shielding fraction of approximately 26%. **b,** Magnetization ( $M$ ) as a function of magnetic field ( $H$ ) at 2 K. The distinct hysteresis loop is characteristic of a type-II superconductor with strong vortex pinning. The lower critical field is estimated to be $\mu_0 H_{c1} \approx 950$ Oe. Using a spherical-sample approximation, the hysteresis width yields a critical current density of $J_c \approx 2{,}000$ A cm$^{-2}$. **c,** Temperature-dependent magnetic torque measured under a field of 0.5 T on an individual $La_6Ni_5O_{12}$ crystal. The torque signal exhibits a sharp anomaly near 18 K, consistent with the superconducting transition observed in susceptibility and resistivity measurements. **Inset**, Magnetic hysteresis loops ( $M_\perp$ vs. $H$ ) measured at 2 K, displaying a pronounced irreversible behavior. **d,** Magnetic hysteresis loops ( $M_\perp$ vs. $H$ ) measured at 5, 10, 15, and 25 K, respectively. The hysteresis width progressively narrows upon warming and vanishes at 25 K, well above the superconducting transition temperature.

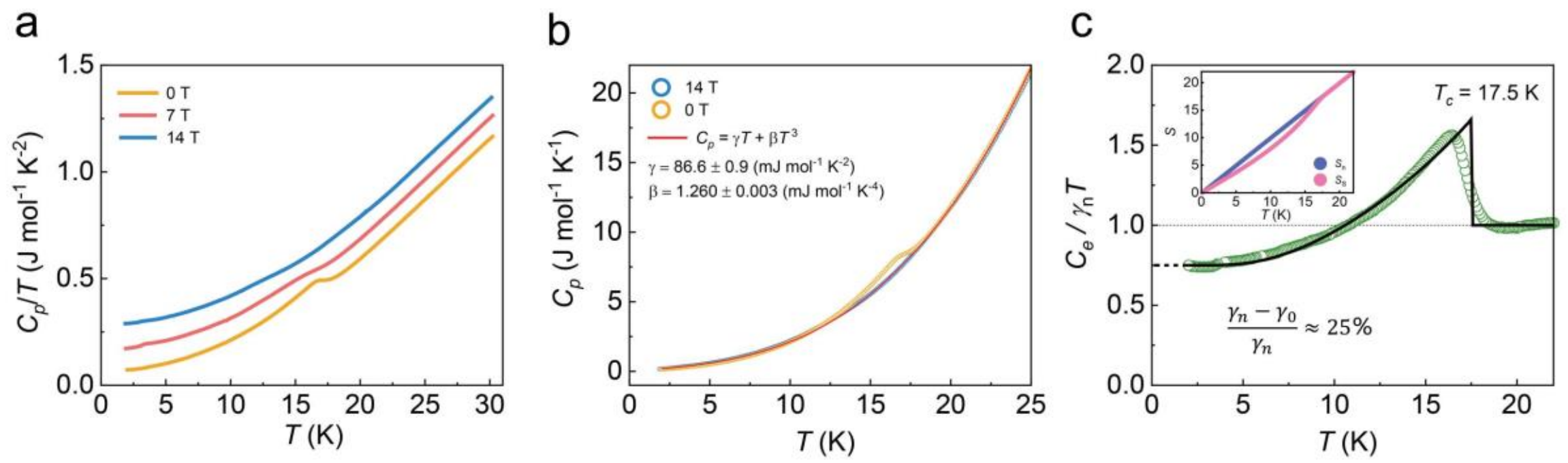

**Fig. 4 | Specific heat analysis and superconducting electronic contribution. a,** $C_p/T$ of polycrystalline $La_6Ni_5O_{12}$ plotted as a function of $T$. Data were taken under $\mu_0 H = 0$ (orange), 7 (pink) and 14 T (blue), respectively; the curves are vertically shifted by 0.1 J mol$^{-1}$ K$^{-2}$ for clarity. **b,** Raw data of $C_p(T)$ measured at 0 T (orange symbols) and 14 T (blue symbols) with a fitting curve adopting the form of $C_p = \gamma_n T + \beta T^3$ (red solid line); the latter was initially fitted to the 14 T data in a window between 14 K and 20 K, yet it satisfyingly traces the zero-$H$ data from $T_c$ up to 25 K. The fit results in $\gamma = 86.6 \pm 0.9$ mJ mol$^{-1}$ K$^{-2}$ and $\beta = 1.260 \pm 0.003$ mJ mol$^{-1}$ K$^{-4}$. **c,** Electronic specific heat $C_e(T)$ obtained after subtracting the lattice contribution determined from the fit in **b**. The data can be well described by the single-band BCS α-model (black solid line) with $\alpha = 2.4$, which gives a mean-field $T_c = 17.5$ K and superconducting fraction of ≈ 25% (Methods). **Inset** shows the entropy balance between the superconducting state ($S_s$) and normal state ($S_n$), see Methods for details.

## Methods

### Synthesis of quintuple-layer nickelate $La_6Ni_5O_{16}$

The quintuple-layer nickelate $La_6Ni_5O_{16}$ were obtained through a two-step procedure. The first step is the preparation of a precursor by a sol-gel route. Specifically, lanthanum nitrate, nickel nitrate, and citric acid were dissolved in deionized water with a molar ratio of La: Ni: citric acid = 5: 4: 15. The resulting solution was heated in an oven at 140 °C for 24 h to evaporate water and obtain a dry gel. Subsequently, the dry gel was calcined in air at 400 °C for 10 h to decompose organic components, yielding well-mixed powder. The as-obtained powder was pressed into pellets and calcined under flowing oxygen at 1100 °C for 10 h to produce the precursor. The phase composition of the precursor was checked by powder XRD which contains a mixture of $La_4Ni_3O_{10}$ and $LaNiO_3$.

The second step is a high-pressure and high-temperature reaction process. In detail, the above precursor were loaded into an alumina crucible, which was further sealed with alumina powder. The sealed crucible was placed in a six-anvil high-pressure high-temperature apparatus for the synthesis of the $La_6Ni_5O_{16}$. The detailed growth parameters were set as follows: The pressure was slowly increased to 5 GPa within 4 h, then the sample was first heated to 1000 °C over 10 min, held at 1000 °C for 30 min, and then further heated to 1600 °C over 30 min. The temperature was held at 1600 °C for 10 h, followed by a slow cooling process down to 1500 °C over 40 h. Afterwards, the sample was quenched to room temperature within 5 h under constant pressure of 5 GPa, and the pressure was released at ambient temperature. The powder XRD patterns were recorded by gridding the whole sample after high temperature and high pressure reaction. Shiny $La_6Ni_5O_{16}$ single crystals could also be isolated from the as-prepared products, with typical dimensions of 0.2 mm × 0.2 mm × 0.06 mm.

### Topotactic reduction with $LiAlH_4$ as the oxygen getter

For the topochemical reduction reaction, approximately 50 mg of the $La_6Ni_5O_{16}$ product was placed in one alumina crucible, while 50 mg of $LiAlH_4$ was loaded into a separate alumina crucible. Both crucibles were sealed inside a quartz tube with an inner diameter of 1.6 cm and a height of 15 cm. The quartz tube was then slowly heated to 310/320/330 °C and held at this temperature for 40 h, respectively. Hydrogen gas was generated via the thermal decomposition of $LiAlH_4$. Upon completion of the reaction, the quartz tube was cracked open in air, and $La_6Ni_5O_{12}$ single crystals could be manually picked out from the resultant product. The powder sample was used to measure the powder XRD data, while the small crystals were used to measure the transport properties.

**Structure and composition characterization**

Single-crystal *X*-ray diffraction data of $La_6Ni_5O_{16}$ and $La_6Ni_5O_{12}$ were collected at room temperature on a four-circle diffractometer (Rigaku, XtaLAB PRO 007HF, Cu *Ka* radiation) at the Life Science Experimental Center, University of Science and Technology of China. The crystal structures were solved and refined using the ShelXT and ShelXL packages implemented in Olex2 software. Detailed crystallographic data are summarized in Extended Data Table 1. Further crystallographic data can be obtained from the CCDC database under deposition numbers 2582907 and 2582925. The powder *X*-ray diffraction patterns were recorded on a Rigaku SmartLab9 diffractometer with Cu *Kα* radiation. We should point out that the powder XRD pattern was collected on material ground from the entire reaction product. The Chemical composition was recorded using a Hitachi SU8220 field emission scanning electron microscope (FE-SEM) equipped with an energy-dispersive *X*-ray spectroscopy (EDS, Oxford Instrument *X*-MaxN 150). The EDS data shown in the Extended Data Fig.4 shows the element ratio of La: Ni close to 1.2:1.

**Scanning transmission electron micrograph (STEM) characterization**

The STEM characterization was conducted at Anhui University. Single-crystal samples were first oriented using a four-circle diffractometer. Subsequently, the specimens were thinned to electron-beam transparency in a dual-beam instrument (Carl Zeiss Crossbeam 550 L FIB-SEM). High-resolution imaging was then performed in a transmission electron microscope (Thermo Fisher Scientific Titan Themis Z microscope) operated at 300 kV. The STEM imags was recorded with a convergence semi-angle of 25 mrad and a collection semi-angle ranging from 8 to 42 mrad for iDPC. The iDPC images were obtained by reconstructing the four segment images, with a high-pass filter applied to suppress the low-frequency contributions.

**Oxygen content determination.**

The average oxygen content of the nickelates was determined by thermogravimetric analysis of the reduction reaction on powder sample: under hydrogen, the nickelate is reduced to $La_2O_3$ and metallic Ni, and the oxygen content is obtained from the accompanying mass loss. In a typical measurement, approximately 50 – 80 mg of powder sample was loaded into a double-walled gold crucible, and its initial mass was precisely recorded with a microbalance. The crucible and 60 mg of $LiAlH_4$ (which decomposes on heating to release hydrogen) were placed in two separate alumina crucibles and sealed in a quartz tube, then heated to 850 °C for 1 h to complete the reduction. After cooling, the gold crucible was reweighed, and the average oxygen content was calculated from the percentage mass change. For as-grown $La_6Ni_5O_{16}$ sample, the determined average oxygen content is 15.9. For $La_6Ni_5O_{12}$ powder samples under different topotactic reduction temperatures, the determined average oxygen contents are 13.1 (310 °C), 12.7 (320 °C), and 12.2 (330 °C), respectively. Because the powder samples may contain minor impurity phases, the average oxygen content obtained from bulk TGA is likely slightly higher than the intrinsic oxygen content of the $La_6Ni_5O_{12}$ phase. More precise determination of the oxygen stoichiometry for the $La_6Ni_5O_{12}$ single crystals will require future microscale characterization, such as multislice electron ptychography [35].

**Resistivity measurement**

Temperature-dependent electrical resistivity was measured using a Quantum Design Physical Property Measurement System (PPMS-14 T) with a standard four-probe alternating-current (AC) technique. Silver paste (DuPont 4929N) was employed and cured at room temperature to fabricate ohmic contacts on the faceted surfaces of single-crystalline $La_6Ni_5O_{16}$ and topotactic-reduced $La_6Ni_5O_{12}$ specimens. For field-dependent resistivity characterizations of $La_6Ni_5O_{12}$, magnetic fields ranging from 0 T to 9 T were applied parallel to the crystallographic $c$-axis, and upper critical field $\mu_0 H_{c2}$ was extracted from the superconducting onset temperatures and fitted via the empirical Ginzburg–Landau model with the formula $H_{c2}(T) = H_{c2}(0)\frac{1-t^2}{1+t^2}$, where $t = T/T_c$.

**Magnetic torque measurement**

Torque magnetometry of $La_6Ni_5O_{12}$ sample was studied using an a piezoresistive self-sensing cantilever probe (SCL Sensor.Tech. PRSA-L300), which was attached to the horizontal rotator of Quantum Design PPMS. A magnetic torque $\boldsymbol{\tau} = \mathbf{M} \times \mathbf{B}$ arises whenever the vector of sample magnetization (**M**) is not collinear with the magnetic inductance vector (**B**). This happens when the sample has anisotropic magnetic susceptibility and **B** is not along any of its high-symmetry axes; as such, torque magnetometry serves as a direct and highly sensitive probe of magnetic anisotropy [40]. In our experiments, the SCL piezoresistive sensor converts the torque exerted on the sample (mounted on the tip of the cantilever) to a voltage signal that can be detected by the measurement of resistance across the circuit incorporated in the sensor. We expected that the change of this resistance reading under magnetic field is proportional to the amplitude of torque in the linear response regime: $\tau \propto R(H) - R(H = 0)$. During the measurement, we fixed the angle between external H-field and the c-axis of small crystal of $La_6Ni_5O_{12}$ to $\theta_{tilt}$ = 45° to maximize the torque signal. Data of $\tau(T)$ were collected under a fixed magnetic field of 0.5 T in a zero-field-cooled process. Isotherms of $\tau(H)$ were measured at temperatures of 2, 5, 10, 15, 25 and 40 K (see Extended Data Fig. 5 and 6). The transverse magnetization $M_\perp$ (the magnetization perpendicular to $H$, rising from the magnetic anisotropy [40]) was calculated from the torque data according to $\boldsymbol{\tau} = \mathbf{M} \times \mu_0 \mathbf{H}$ (which is valid as long as $M \ll H$), i.e., $M_\perp = \tau/\mu_0 H$. The sign of $M_\perp$ is determined with the reference of the high-$H$ part of $\tau(H)$ in the normal state ($T$ = 40 K), which must be positive (paramagnetic).

**Magnetic susceptibility measurement and the estimation of the superconducting critical current density**

DC magnetic susceptibility and magnetization isotherms were acquired on the Quantum Design MPMS5 system. Zero-field-cooled (ZFC) and field-cooled (FC) susceptibility curves were recorded under a small applied field of 100 Oe over the temperature range of 2–40 K. Magnetization–field ($M$-$H$) hysteresis loops were collected at $T$ = 2 K within ±0.5 T to determine the lower critical field $\mu_0 H_{c1}$ and verifies the superconducting nature of tetragonal $La_6Ni_5O_{12}$ single crystals.

The superconducting critical current density, $J_c$, was estimated from the magnetic hysteresis loop measured at 2 K using the Bean critical-state model. For a sphere sample, the $J_c$ can be estimated with the formula $J_c = \frac{5\Delta M(H)}{2R}$. The characteristic radius $R$ was determined based on the total mass. The irreversible magnetization width, $\Delta M(H)$, was obtained from the difference between the magnetization values measured during decreasing- and increasing-field sweeps close to the zero magnetic field. $J_c$ was then calculated from $\Delta M$ using the spherical-sample form of the critical-state model. The resulting self-field critical current density at 2 K is approximately 2000 A $cm^{-2}$.

**Specific heat measurement**

Specific heat was measured using a relaxation-time method in a Dynacool-14T physical property measurement system (Quantum Design, PPMS). Polycrystalline $La_6Ni_5O_{12}$ sample of mass 13 mg was pressed into a pellet and used in this experiment. To separate the contributions from electrons ($C_e$) and phonons ($C_{ph}$), we fitted the normal-state data by $C_p(T)/T = \gamma_n + \beta T^2$. This fit was performed for both the zero-$H$ and the 14 T data and they yielded consistent results (Fig. 4b): both can be well fitted using $\gamma_n$ = 86.6 mJ mol$^{-1}$ K$^{-2}$ and $\beta$ = 1.260 mJ mol$^{-1}$ K$^{-4}$ (with reasonable uncertainties, as shown in Fig. 4b). We then subtracted $C_{ph} = \beta T^3$ from the zero-$H$ $C_p(T)$ to extract the $C_e$ term in the superconducting state (see Fig. 4c). Here the $C_e/T$ includes the contribution from the non-superconducting composition(s) in the samples, which we treated as a normal metal with a Sommerfeld coefficient $\gamma = \gamma_0$ and accounted for it in a fitting based on BCS $\alpha$-model [37, 38] (black curve in Fig. 4c). $C_e(T)/T$ data can be satisfyingly fitted by this $\alpha$-model assuming $T_c$ = 17.5 K and single gap $\Delta$ = 3.6 meV, which corresponds to $2\Delta/k_BT_c$ = 4.8, a ratio apparently exceeding the weak-coupling BCS values (3.53). Hence, the superconductivity in $La_6Ni_5O_{12}$ is more likely to be strong-coupling. Furthermore, assuming that the residual $C_e/T$ at $T$ = 0 is fully attributed to the $\gamma_0$ term, we can inter $\gamma_0/\gamma_n \approx 75\%$. (Note that we calculated the molar specific heat according to the molar mass and unit-cell volume of $La_6Ni_5O_{12}$, hence the estimation may be subject to a certain degree of uncertainty; this notwithstanding, it ought to be qualitatively valid.) As such, the superconducting phase contributes to $1 - \gamma_0/\gamma_n \approx 25\%$ of the total electronic specific heat. We further point out that our estimation of superconducting phase fraction (and the background subtraction) is validated by the analysis of entropy balance [39]. Here, the entropy $S_s(T)$ of the superconducting state was calculated by integrating the extracted $C_e/T$ from zero to $T$; for $T < 2$ K where the data are unavailable, we exploited the extrapolation of the $\alpha$-model fit down to $T$ = 0, see the dashed line in Fig. 4c. The normal state entropy $S_n(T)$ was estimated based on the fitting curve in Fig. 4b. As shown in the inset of Fig. 4c, $S_s(T)$ demonstrates a recovery towards the value of the normal-state entropy $S_n(T)$ at a temperature (17.8 K) close to $T_c$ = 17.5 K yielded by the $\alpha$-model fit. Hence, the entropy balance condition is satisfied, implying that the $C_e/T$ data shown in Fig. 4c reflect the contribution of superconducting phase reasonably well.

## Reference

1. Bednorz, J. G. & Muller, K. A. Possible high $T_c$ superconductivity in the Ba–La–Cu–O system. *Z. Phys. B: Condens. Matter* **64**, 189-193 (1986).
2. Keimer, B., Kivelson, S. A., Norman, M. R., Uchida, S. & Zaanen, J. From quantum matter to high-temperature superconductivity in copper oxides. *Nature* **518**, 179–186 (2015).
3. Norman, M. R. Materials design for new superconductors. *Rep. Prog. Phys.* **79**, 074502 (2016).
4. Anisimov, V. I., Bukhvalov, D. & Rice, T. M. Electronic structure of possible nickelate analogs to the cuprates. *Phys. Rev. B* **59**, 7901-7906 (1999).
5. Li, D. F. *et al.* Superconductivity in an infinite-layer nickelate. *Nature* **572**, 624-627 (2019).
6. Sahib, H. *et al.* Superconductivity in $PrNiO_2$ Infinite-Layer Nickelates. *Adv. Mater.* **37**, 2416187 (2025).
7. Parzyck, C. T. *et al.* Superconductivity in the Parent Infinite-Layer Nickelate $NdNiO_2$. *Phys. Rev. X* **15**, 021048 (2025).
8. Chow, S. Lin Er *et al.* Bulk superconductivity near 40 K in hole-doped $SmNiO_2$ at ambient pressure. *Nature* **642**, 58–63 (2025).
9. Pan, G. A. *et al.* Superconductivity in a quintuple-layer square-planar nickelate. *Nat. Mater.* **21**, 160-164 (2022).
10. Sun, H. *et al.* Signatures of superconductivity near 80 K in a nickelate under high pressure. *Nature* **621**, 493–498 (2023).
11. Wang, N. *et al.* Bulk high-temperature superconductivity in pressurized tetragonal $La_2PrNi_2O_7$. *Nature* **634**, 579–584 (2024).
12. Zhu, Y. *et al.* Superconductivity in pressurized trilayer $La_4Ni_3O_{10-\delta}$ single crystals. *Nature* **631**, 531–536 (2024).
13. Shi, M. *et al.* Pressure induced superconductivity in hybrid Ruddlesden–Popper $La_5Ni_3O_{11}$ single crystals. *Nat. Phys.* **21**, 1780–1786 (2025).
14. Yan, X. *et al.* Superconductivity in an ultrathin multilayer nickelate. *Sci. Adv.* **11**, eado4572 (2025).

15. Ko, E. K. *et al.* Signatures of ambient pressure superconductivity in thin film $La_3Ni_2O_7$. *Nature* **638**, 935–940 (2025).

16. Zhou, G. *et al.* Ambient-pressure superconductivity onset above 40 K in $(La,Pr)_3Ni_2O_7$ films. *Nature* **640**, 641–646 (2025).

17. Li, Q. *et al.* Absence of superconductivity in bulk $Nd_{1-x}Sr_xNiO_2$. *Commun. Mater.* **1**, 16 (2020).

18. Wang, B.-X. *et al.* Synthesis and characterization of bulk $Nd_{1-x}Sr_xNiO_2$ and $Nd_{1-x}Sr_xNiO_3$. *Phys. Rev. Mater.* **4**, 084409 (2020).

19. Puphal, P. *et al.* Topotactic transformation of single crystals: From perovskite to infinite-layer nickelates. *Sci. Adv.* **7**, eabl8091 (2021).

20. Puphal, P. *et al.* Phase formation in hole- and electron-doped rare-earth nickelate single crystals. *APL Mater.* **11**, 081107 (2023).

21. Suyolcu, Y. *et al.* Three generations of infinite-layer nickelate crystals. *MRS Communications* **15**, 169–180 (2025).

22. Hu, K. *et al.* Atomic scale disorder and reconstruction in bulk infinite-layer nickelates lacking superconductivity. *Nat. Commun.* **15**, 5104 (2024).

23. Lacorre, Ph. Passage from T-type to T′-type arrangement by reducing $R_4Ni_3O_{10}$ to $R_4Ni_3O_8$ (R = La, Pr, Nd). *J. Solid State Chem.* **97**, 495–500 (1992).

24. Zhang, Y. *et al.* Large orbital polarization in a metallic square-planar nickelate. *Nat. Phys.* **13**, 864 (2016).

25. Ferenc Segedin, D. *et al.* Limits to the strain engineering of layered square-planar nickelate thin films. *Nat. Commun.* **14**, 1468 (2023).

26. LaBollita, H. & Botana, A. S. Correlated electronic structure of a quintuple-layer nickelate. *Phys. Rev. B* **105**, 085118 (2022).

27. Jung, M. C., Kapeghian, J., Hanson, C., Pamuk, B. & Botana, A. S. Electronic structure of higher-order Ruddlesden-Popper nickelates. *Phys. Rev. B* **105**, 085150 (2022).

28. LaBollita, H. & Botana, A. S. Electronic structure and magnetic properties of higher-order layered nickelates: $La_{n+1}Ni_nO_{2n+2}$ (n=4-6). *Phys. Rev. B* **104**, 035148 (2021).

29. Carrasco-Alvarez, A. A. *et al.* Electron-phonon mediated superconductivity in $La_6Ni_5O_{12}$ nickel

oxides. *Phys. Rev. B* **110**, 235107 (2024).

30. Pan, G. A. *et al.* Superconducting phase diagram of multilayer square-planar nickelates. *Science* **392**, 1390-1395 (2026).

31. Poltavets, V. V. *et al.* $La_3Ni_2O_6$: A new double T′-type nickelate with infinite $Ni^{1+/2+}$ $O_2$ layers. *J. Am. Chem. Soc.* **128**, 9050-9051 (2006).

32. Poltavets, V. V. *et al.* Crystal structures of $Ln_4Ni_3O_8$ (Ln = La, Nd) triple layer T′-type nickelates. *Inorg. Chem.* **46**, 10887-10891 (2007).

33. Zhao, D. *et al.* Pressure-enhanced spin-density-wave transition in double-layer nickelate $La_3Ni_2O_{7-\delta}$. *Sci. Bull.* **70**, 1239-1245 (2025).

34. Wang, Y. *et al.* Unconventional density-wave state in Ruddlesden–Popper nickelate $La_4Ni_3O_{10}$. *Nat. Commun.* **17**, 6422 (2026).

35. Dong, Z. *et al.* Visualization of oxygen vacancies and self-doped ligand holes in $La_3Ni_2O_{7-\delta}$. *Nature* **630**, 847–852 (2024).

36. Liu, Y. D. *et al.* Superconductivity and normal-state transport in compressively strained $La_2PrNi_2O_7$ thin films. *Nat. Mater.* **24**, 1221–1227 (2025).

37. Padamsee, H. *et al.* Quasiparticle Phenomenology for Thermodynamics of Strong-Coupling Superconductors. *J. Low Temp. Phys.* **12**, 387 (1973).

38. Johnston, D. C. Elaboration of the α-model derived from the BCS theory of superconductivity, *Supercond. Sci. Technol.* **26**, 115011 (2013).

39. Tallon, J. L., Storey, J. G. & Loram, J. W. Fluctuations and critical temperature reduction in cuprate superconductors. *Phys. Rev. B* **83**, 092502 (2011).

40. Li, L. Torque Magnetometry in Unconventional Superconductors. Ph.D. Thesis, Princeton University (2008).

## Acknowledgements

We acknowledge fruitful discussions with Zhenyu Wang, and the experimental assistance from Jianjun Ying, Binghui Ge, Feng Cheng and Zhongliang Zhu. This work is supported by the National Natural Science Foundation of China (Grant Nos. 12488201, 12494592, 12325403, 12574149), the National Key R&D Program of the MOST of China (Grant No. 2022YFA1602601), the Chinese Academy of Sciences under contract No. JZHKYPT-2021-08, the CAS Project for Young Scientists in Basic Research (Grant No. YBR-048), the CAS Superconducting Research Project under Grant No. [SCZX-0101] and the Innovation Program for Quantum Science and Technology (Grant No. 2021ZD0302800).

## Author Contributions

X.H.C. conceived the research project and coordinated the experiments. M.Z.S. grew the single crystals, and performed the structural and magnetic susceptibility characterization. M.D. measured the magnetic torque data and specific heat with the help of H.Y.L. and Z.J.X., and measured the resistance data with Z.W.W. and Z.J.W.  M.Z.S. and M.D. measured the oxygen content by TGA. M. D., M.Z.S., Z.J.X., T.W. and X.H.C. analyzed the data. M.Z.S., Z.J.X., T.W. and X.H.C. wrote the paper with input from all the authors.

## Competing Interests

The authors declare no competing interests.